\documentclass[iop,revtex4]{emulateapj}

\usepackage{txfonts}
\usepackage{textcomp}

\shorttitle{Low mass Ly$\alpha$ emitter at $z$=5.75}
\shortauthors{Hern\'an-Caballero et al.}
\begin{document}
\title{SHARDS Frontier Fields: physical properties of a low mass Lyman-$\alpha$ emitter at \MakeLowercase{z}=5.75}
\author{
Antonio Hern\'an-Caballero,\altaffilmark{1}
Pablo G. P\'erez-Gonz\'alez,\altaffilmark{1} 
Jose M. Diego,\altaffilmark{2} 
David Lagattuta,\altaffilmark{3}
Johan Richard,\altaffilmark{3} 
Daniel Schaerer,\altaffilmark{4} 
Almudena Alonso-Herrero,\altaffilmark{5} 
Raffaella Anna Marino,\altaffilmark{6}
Panos Sklias,\altaffilmark{1} 
Bel\'en Alcalde-Pampliega,\altaffilmark{1} 
Antonio Cava,\altaffilmark{4}
Christopher J. Conselice,\altaffilmark{7}
Helmut Dannerbauer,\altaffilmark{8,9}
Helena Dom\'inguez-S\'anchez,\altaffilmark{10} 
Carmen Eliche-Moral,\altaffilmark{8,9} 
Pilar Esquej,\altaffilmark{11} 
Marc Huertas-Company,\altaffilmark{10,12,13}
Rui Marques-Chaves,\altaffilmark{8,9}
Ismael P\'erez-Fournon,\altaffilmark{8,9}
Tim Rawle,\altaffilmark{14}
Jos\'e Miguel Rodr\'iguez Espinosa,\altaffilmark{8,9} 
Daniel Rosa Gonz\'alez,\altaffilmark{15}
Wiphu Rujopakarn\altaffilmark{16}
} 
\email{a.hernan@ucm.es}
\altaffiltext{1}{Departamento de Astrof\'isica y CC. de la Atm\'osfera, Facultad de CC. F\'isicas, Universidad Complutense de Madrid, E-28040 Madrid, Spain}
\altaffiltext{2}{Instituto de F\'isica de Cantabria, CSIC-UC, Avenida de los Castros s/n, 39005, Santander, Spain}
\altaffiltext{3}{Univ Lyon, Univ Lyon1, Ens de Lyon, CNRS, Centre de Recherche Astrophysique de Lyon UMR5574, F-69230, Saint-Genis-Laval, France}
\altaffiltext{4}{Observatoire de Geneve, Univ Geneve, 51, Ch. des Maillettes, CH-1290 Versoix, Switzerland}
\altaffiltext{5}{Centro de Astrobiolog\'ia (CSIC-INTA), ESAC Campus, E-28692 Villanueva de la Ca\~nada, Madrid, Spain}
\altaffiltext{6}{Institute for Astronomy, Department of Physics, ETH Z\"urich, Switzerland}
\altaffiltext{7}{School of Physics and Astronomy,University of Nottingham,
University Park, NG9 2RD, UK}
\altaffiltext{8}{Instituto de Astrof\'\i sica de Canarias, C/V\'\i a L\'actea, s/n, E-38205 San Crist\'obal de La Laguna, Tenerife, Spain}
\altaffiltext{9}{Universidad de La Laguna, Dpto. Astrof\'\i sica, E-38206 La Laguna, Tenerife, Spain}
\altaffiltext{10}{LERMA, Observatoire de Paris, PSL Research University, CNRS, Sorbonne Universités, UPMC Univ. Paris 06, F-75014, Paris, France}
\altaffiltext{11}{European Space Astronomy Centre (ESAC)/ESA, E-28691 Villanueva de la Ca\~nada, Madrid, Spain}
\altaffiltext{12}{Universit\'e Paris Diderot, 5 Rue Thomas Mann, 75013, France}
\altaffiltext{13}{Department of Physics and Astronomy, University of Pennsylvania, Philadelphia, PA 19104, USA}
\altaffiltext{14}{ESA / Space Telescope Science Institute (STScI), 3700 San Martin Drive, Baltimore, MD 21218, USA}
\altaffiltext{15}{Instituto Nacional de Astrof\'isica, \'Optica y Electr\'onica, Puebla, Mexico}
\altaffiltext{16}{Kavli Institute for the Physics and Mathematics of the Universe (WPI), University of Tokyo Institutes for Advanced Study,
Kashiwa, Chiba 277-8583, Japan}

\begin{abstract}
We analyze the properties of a multiply-imaged Lyman-$\alpha$ (Ly$\alpha$) emitter at $z$=5.75 identified through SHARDS Frontier Fields intermediate-band imaging of the Hubble Frontier Fields (HFF) cluster Abell 370. The source, A370-L57, has low intrinsic luminosity (M$_{UV}$$\sim$-16.5), steep UV spectral index ($\beta$=-2.4$\pm$0.1), and extreme rest-frame equivalent width of Ly$\alpha$ (EW$_0$(Ly$\alpha$)=420$^{+180}_{-120}$ \AA). Two different gravitational lens models predict high magnification ($\mu$$\sim$10--16) for the two detected counter-images, separated by 7'', while a predicted third counter-image ($\mu$$\sim$3--4) is undetected. 
We find differences of $\sim$50\% in magnification between the two lens
models, quantifying our current systematic uncertainties.
Integral field spectroscopy of A370-L57 with MUSE shows a narrow ($FWHM$=204$\pm$10 km s$^{-1}$) and asymmetric Ly$\alpha$ profile with an integrated luminosity L(Ly$\alpha$)$\sim$10$^{42}$ erg s$^{-1}$.
The morphology in the HST bands comprises a compact clump ($r_e$$<$100 pc) that dominates the Ly$\alpha$ and continuum emission and several fainter clumps at projected distances $\lesssim$1 kpc that coincide with an extension of the Ly$\alpha$ emission in the SHARDS F823W17 and MUSE observations. The latter could be part of the same galaxy or an interacting companion. We find no evidence of contribution from AGN to the Ly$\alpha$ emission. 
Fitting of the spectral energy distribution with stellar population models favors a very young ($t$$<$10 Myr), low mass (M$_*$$\sim$10$^{6.5}$ M$_\odot$), and metal poor ($Z$$\lesssim$4$\times$10$^{-3}$) stellar population. 
Its modest star formation rate (SFR$\sim$1.0 M$_\odot$ yr$^{-1}$) implies high specific SFR ($sSFR$$\sim$2.5$\times$10$^{-7}$ yr$^{-1}$) and SFR density ($\Sigma_{SFR}$$\sim$7--35 M$_\odot$ yr$^{-1}$ kpc$^{-2}$). 
The properties of A370-L57 make it a good representative of the population of galaxies responsible for cosmic reionization. 

\end{abstract}


\section{Introduction} 

The recombination that generated the cosmic microwave background (CMB) occurred at $z$$\sim$1000. Subsequently the gas content of the Universe remained neutral (in what is known as the cosmic dark ages) until the first stars started the process of reionization.
The latest measurements of the optical depth to electron scattering in the CMB indicate a redshift for reionization (assumed instantaneous) of $z$=8.8$^{+1.7}_{-1.4}$ \citep{PlanckCollaboration16}. However, reionization was an extended process that likely started at $z$$>$10 \citep{Stark10,Ono12,Robertson15} and was completed only by $z$$\sim$6 \citep{Fan06,Mortlock11}. 

While active galactic nuclei (AGN)  may have contributed a non-negligible fraction of the required ionizing photons \citep{Giallongo15,Madau15}, the prevailing view is that reionization was largely driven by the rate of escape into the intergalactic medium (IGM) of Lyman continuum (LyC) photons from star-forming galaxies \citep[e.g.][]{Stiavelli04,Richards06,Robertson10,Bouwens15a}.
The details of this process are, however, poorly constrained by current observations due to large uncertainties in the derived escape fraction of LyC photons and the star formation rate density at high redshifts.

The UV continuum luminosity function (LF) of high redshift galaxies is now well constrained up to $z$$\sim$8 \citep[e.g.][]{Atek15,Bouwens15b,Finkelstein15}. It shows significant evolution, with an increasingly steeper slope at higher redshift \citep{Alavi14,Bouwens15b,Livermore17}. Together with an increased escape probability for LyC photons in less massive galaxies \citep{Dijkstra16,Faisst16}, this implies that the ionizing photon budget was dominated by the much more numerous galaxies at the faint end of the UV LF.
Therefore, quantifying and characterizing this population are essential to our understanding of reionization. Furthermore, in the current paradigm of hierarchical galaxy assembly, these early low mass galaxies are believed to be the building blocks that merged to form the L* galaxies seen at lower redshifts \citep[e.g.][]{Dressler11}.

Known galaxies at high redshift belong to one of two classes that differ in the way they are selected: Lyman Break Galaxies (LBGs) and Lyman-$\alpha$ Emitters (LAEs). Classically, LBGs are identified in broadband surveys by the break in the continuum at 912 \AA{} due to the Lyman limit \citep{Steidel96}. However, at $z$$\gtrsim$5 the Ly$\alpha$ forest caused by intervening neutral hydrogen clouds becomes so dense in lines that the 912--1216 \AA{} continuum is strongly depleted, forming a new break at the wavelength of Ly$\alpha$ (1216 \AA) more prominent than the actual Lyman break. 
LAEs, on the other hand, are galaxies identified by their strong Ly$\alpha$ emission.
While all star-forming galaxies produce copious amounts of Ly$\alpha$ photons in their HII regions, the high cross-section of neutral hydrogen to Ly$\alpha$ photons implies that they are absorbed and re-emitted in random directions multiple times before they can escape to the IGM, increasing the probability of absorption by dust grains. As a consequence, many LBGs do not show Ly$\alpha$ emission, and galaxies with strong Ly$\alpha$ emission tend to be less massive and contain less dust compared to LBGs \citep{Giavalisco02,Gawiser07}.

Because Ly$\alpha$ can be very bright compared to the UV continuum, searching for LAEs is the most efficient method to find the least massive galaxies. 
In the last decade, large area narrow band surveys have identified hundreds of LAEs at $z$$\sim$5.7 \citep[e.g.][]{Ouchi08,Santos16,Ouchi17} and $z$$\sim$6.6 \citep{Ouchi10,Santos16,Ouchi17}, and dozens at $z$$\sim$7-8 \citep{Ota10,Shibuya12,Konno14,Ota17,Zheng17}.
However, as we advance into the reionization epoch, the neutral gas fraction increases, and while luminous LAEs capable of ionizing large bubbles have been observed up to $z$$\sim$8.7 \citep{Zitrin15}, the Ly$\alpha$ emission of low luminosity LAEs diffuses in the surrounding neutral hydrogen, becoming unobservable. This is evidenced by the steep decline at $z$$>$6 in the number density of faint LAEs detected by narrow band surveys \citep{Ouchi10,Konno14,Santos16}, the marked decline in the average equivalent width (EW) of Ly$\alpha$ in continuum selected galaxies \citep{Fontana10,Stark10,Ono12,Schenker12,Schenker14}, and the increase in the size of Ly$\alpha$ haloes \citep{Santos16}. 
Therefore faint LAEs at $z$$\sim$6 may currently be the best proxies to infer the properties of the galaxies that reionized the Universe \citep{Dawson13}.

The Ly$\alpha$ luminosity function of LAEs at z$\sim$6 is now well constrained down to L$_{Ly\alpha}$ $\sim$ 10$^{42.5}$ erg s$^{-1}$ \citep{Dressler15,Santos16}.
However, observational limitations imply that detailed studies of LAEs at these redshifts have largely been limited to the most luminous ones.
For the sub-L* LAEs at $z$$\sim$6, often the only information available is their Ly$\alpha$ luminosity, and -if there are deep enough continuum observations- a rough estimate of the Ly$\alpha$ EW. 
A significant fraction of these LAEs \citep[10-40\% at z=5.7 according to][]{Shimasaku06} have rest-frame Ly$\alpha$ EW$>$240 \AA, which cannot be reproduced with the stellar population models commonly used for lower redshift galaxies. Instead, they require a top heavy initial mass function (IMF), very low metallicity, and/or very young ($<$10$^{7}$ Myr) ages \citep{Charlot93,Malhotra02}. 

Our best chance at studying this population in detail with current technology is to identify faint LAEs magnified by the strong lensing effect of a galaxy cluster \citep[e.g.][]{Ellis01,Santos04,Richard11}. 
The Hubble Frontier Fields \citep[HFF;][]{Lotz17} recently obtained Advanced Camera for Surveys (ACS) and Wide Field Camera 3 (WFC3) imaging of six galaxy clusters to a depth of $\sim$29 mag AB, only matched by the Hubble Ultra Deep Field. Integral field spectroscopy of HFF clusters with the Multi Unit Spectroscopic Explorer \citep[MUSE;][]{Bacon10} on the Very Large Telescope has revealed several dozen lensed LAEs at 3$<$$z$$<$6 \citep{Caminha17,Karman17,Lagattuta17,Mahler17}.
Two of these clusters, Abell 370 and MACS J1149.5+2223, are also being observed by the SHARDS Frontier Fields survey (SHARDS-FF; PI: P\'erez-Gonz\'alez), an imaging survey with the Gran Telescopio Canarias (GTC) that covers the 500--950 nm spectral range with 25 medium band filters (R$\sim$50) down to m$_{AB}$$\sim$27.
The SHARDS-FF observations allow us to select LAEs fainter than L* (even without magnification) at redshift up to $z$$\sim$7. 

In this paper we present the analysis of physical properties of the first faint LAE identified through SHARDS-FF observations. The source, A370-L57, is a triply imaged galaxy at $z$=5.75 lensed by the Abell 370 galaxy cluster. Its magnification corrected UV luminosity (M$_{UV}\sim$-16.5) is comparable to the faintest LAEs identified in other Frontier Fields clusters \citep[e.g.][]{Karman17,Caminha17,Vanzella17}, while its extreme Ly$\alpha$ EW (EW$_0$(Ly$\alpha$)=420$^{+180}_{-120}$ \AA) is the largest yet found for any LAE in the Frontier Fields.

The paper is structured as follows: in \S2 we describe the observations and data reduction process, while \S3 presents the two lens models for Abell 370 that we have considered. In \S4 we outline our method for selection of emission line galaxies on the SHARDS-FF images. \S5 characterizes the A370-L57 galaxy in terms of apparent morphology of the stellar and nebular emission, the spectral energy distribution, and the properties of the Ly$\alpha$ emission line. 
In \S6 we determine magnification-corrected values for the UV continuum and Ly$\alpha$ luminosity, star formation rate (SFR), and effective radius. In \S7 we use stellar population models to fit the spectral energy distribution (SED) and estimate the age, metallicity, and mass of the young stellar population in the galaxy. \S8 discusses the possibility of AGN and low-metallicity star formation in the galaxy. Finally, \S9 summarizes our main conclusions.

Throughout this paper we assume a $\Lambda$CDM cosmology with $\Omega_\Lambda$ = 0.714, $\Omega_M$ = 0.286, H$_0$ = 69.6 km s$^{-1}$ Mpc$^{-1}$. All magnitudes are in the AB system.  

\section{Observations and data}\label{sec:observations}

The SHARDS Frontier Fields (P\'erez-Gonz\'alez et al. in preparation) is an ongoing long-term observational program with GTC/OSIRIS. The targets are two of the HFF galaxy clusters: Abell 370 and MACS J1149.5+2223. The 8.5'$\times$7.8' FOV of OSIRIS covers both the main and parallel HST Hubble Frontier Fields observations with a single pointing. A total of 240 hours have been assigned to the program. The observations started in December 2015, and upon completion, they will reach at least 3-$\sigma$ sensitivity of m$\sim$27 in all 25 medium band filters.
The full depth of the SHARDS-FF survey was achieved first for the F823W17 filter
($\lambda_{eff}$=823 nm, FWHM=14.7 nm) in the Abell 370 field. 34 exposures were taken between Dec 2015 and Jan 2016 totaling 5.28 hours of integration time. 

We reduced the individual images in the F823W17 band using our custom OSIRIS pipeline described in \citet{Perez-Gonzalez13}. In addition to bias subtraction and flat fielding, it includes illumination correction, background gradient subtraction, fringing removal, World Coordinate System (WCS) alignment including field distortions, two-dimensional calibration of the passband and zero point, and, finally, stacking of the individual frames. 
The final F823W17 image has a pixel scale of 0.25'' and the PSF FWHM is 0.78''. The limiting magnitude at 5-$\sigma$ is m=26.8.

We retrieved from the Mikulski Archive for Space Telescope\footnote{https://archive.stsci.edu/prepds/frontier/} (MAST) the reduced public mosaics from the v1.0 release of Epochs 1 and 2, which combine all the HFF observations of Abell 370 as well as data from previous imaging programs, in the ACS filters F435W, F606W, and F814W, and the WFC3 filters F105W, F125W, F140W, and F160W. The limiting magnitude is $\sim$29 (5-$\sigma$) in all the bands.
Among the several available flavors of the mosaics we chose the ones with 0.03'' pixel scale and processed with the self-calibration option (in the case of ACS data) and the time-variable sky background subtraction option (in the case of WFC3 data).

The central region of the Abell 370 cluster has also been targeted for integral field spectroscopy with MUSE by the GTO program 094.A-0115A (PI: Richard) and the GO program 096.A-0710A (PI: Bauer). 

We retrieved from the ESO archive the fully reduced data cubes (PHASE 3) for the MUSE observations that are already public by the time of this writing (April 2017). The data were reduced with version 1.6.1 of the MUSE Instrument Pipeline.
The pixel scale of the data cubes is 0.2'' and the spectral resolving power is R = $\lambda$/$\delta\lambda$ = 3026. The exposure times range from 2700 to 3450 seconds, and the PSF FWHM ranges from 0.6'' to 0.8''.
We adjusted the astrometry of the data cubes by extracting a synthetic image using the transmission curve of the F814W filter and aligning to the F814W image from HFF. We checked the absolute flux calibration of the data cubes by comparing synthetic photometry in the F814W and F823W17 filters with that from the HFF and SHARDS-FF images. For sources with S/N$>$10 the dispersion is $\sim$10\% with no significant bias. 

\section{Lens models}\label{sec:lensmodels}

We estimate the magnification and shear distortion of background galaxies by the gravitational lens using the mass models for Abell 370 generated by \citet[][hereafter D16]{Diego16} and \citet[][L17]{Lagattuta17}. 

The D16 model uses a free-form mass distribution that, starting from a reliable set of 10 multiply lensed systems, identifies $\sim$80 multiple-images. The lensing mass reconstruction is performed with the WSLAP+ method \citep{Diego05,Diego07,Diego16,Sendra14}.

The mass in the lens plane is modeled as a combination of a diffuse component and a compact component. The diffuse component is a superposition of Gaussian functions located at a distribution of grid points which can be regular or adaptive. The compact mass accounts for the mass (baryonic and dark matter) associated with the member galaxies. Usually the distribution of mass is assumed to follow the distribution of light for the compact component. The member galaxies are selected from the red sequence and are elliptical-type galaxies.

The L17 model describes the mass of the cluster as a distribution of clumps, including large scale dark-matter halos representing cluster potentials and smaller galaxy-scale halos representing individual galaxies. Each halo is assumed to have a truncated Dual Pseudo-Isothermal Elliptical mass distribution \citep[dPIE;][]{Eliasdottir07}. The model parameters and their uncertainties are computed with the LENSTOOL software \citep{Kneib96,Jullo07,Jullo09}.

While the selection criteria for cluster members are not identical, differences only affect small galaxies whose impact in the lens model is negligible.
The parameters of both mass models are adjusted using the systems of multiple images identified in the HFF images, some of them with spectroscopic redshifts from the Grism Lens-Amplified Survey from Space \citep[GLASS;][]{Treu15} and previous spectroscopic campaigns \citep{Richard10,Richard14,Johnson14}. L17 also adds spectroscopic redshifts for 10 multiple image systems from the MUSE observations of the program 094.A-0115A. The number of systems with spectroscopic redshift, the total number of systems, and the total number of images are 7/30/83 and 17/22/69 for D16 and L17, respectively.

In the D16 model, the typical RMS between the observed and predicted positions is $\sim$1'' for well constrained systems and a few arcseconds for less constrained systems (i.e systems with no known spectroscopic redshift and/or in a region of the lens plane with no additional lensing constraints). 
RMS values in the L17 model range between 0.2'' and 1.5''. The total model RMS is 0.94''.

\section{Identification of a \MakeLowercase{z}=5.75 Ly$\alpha$ emitter}

We searched for emission line galaxies in the central region of Abell 370 by subtracting a PSF-matched version of the F814W image from the SHARDS F823W17 image. We do this with a custom optimal image subtraction routine that performs PSF-matching using a spatially varying analytical kernel \citep[see e.g.][]{Alard98,Alard00}. This routine is described in detail in an upcoming paper Hern\'an-Caballero et al. (in preparation). 
Very briefly, the workflow is arranged in four stages: a) removal of the residual sky background in both images; b) resampling of the F814W image to match the pixel layout of the F823W17 image; c) convolution of the F814W image with a kernel that varies through the image, to compensate for the spatial variation of the PSF in both the F814W and F823W17 images; d) fine-tuning the absolute flux calibration of the F823W17 image using F814W as reference. 
Steps c) and d) are performed iteratively. In each iteration the convolved F814W image is subtracted from the flux-calibrated F823W17 image and the mean absolute residual (\textit{mar}) is evaluated. The parameters of the analytical function that defines the convolution kernel are adjusted in order to minimize the \textit{mar}. 
The result is a `residual image' where sources with a color index [F823W17]-[F814W]$\sim$0 are largely removed, sources with [F823W17]-[F814W]$>$0 appear as negative residuals, and those with [F823W17]-[F814W]$<$0 (the expected outcome of an emission line affecting the F823W17 flux) appear as positive residuals. 

\begin{figure*} 
\begin{tabular}{p{0.6\textwidth} p{0.02\textwidth} p{0.35\textwidth}}
\vspace{0pt}
\hspace{-0.05cm}\includegraphics[height=9.95cm]{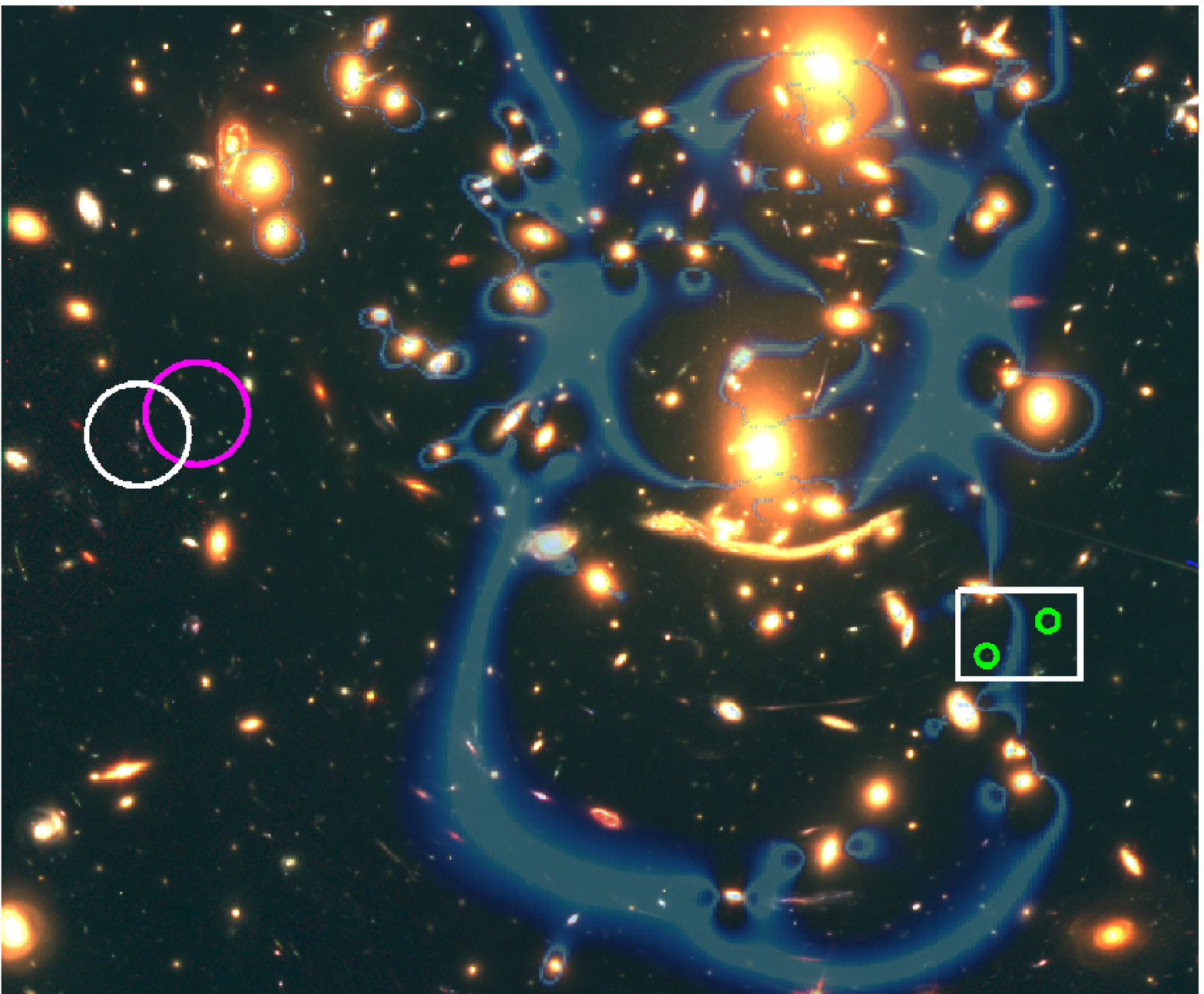} & &
\vspace{-0.23cm}\includegraphics[height=10.45cm]{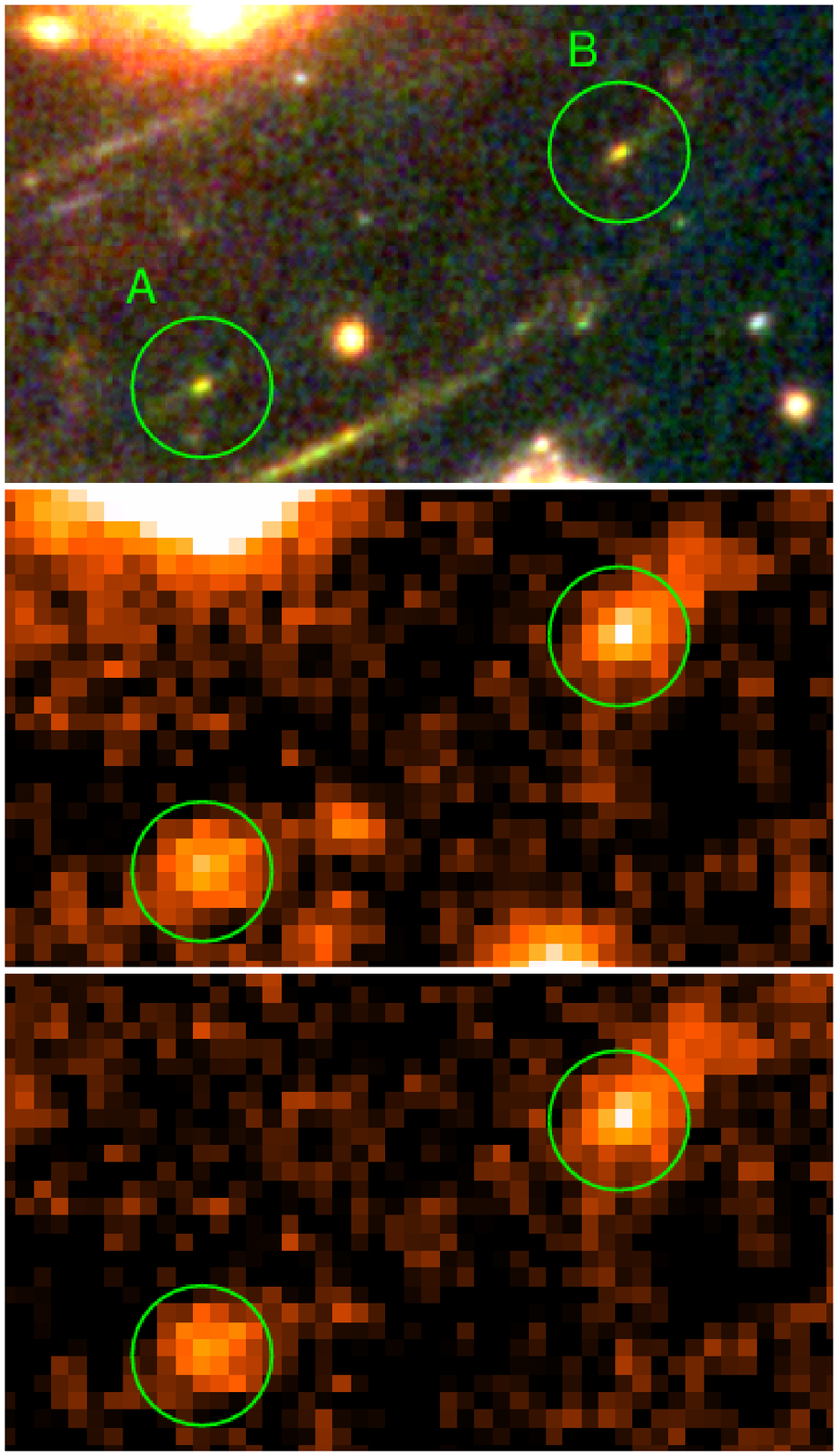}
\end{tabular}
\caption[]{Left: RGB composite image of a 115''$\times$100'' region in the central part of the Abell 370 galaxy cluster, obtained by combining HFF frames in the F160W (red), F814W (green), and F435+F606W (blue) filters. The cyan overlay represents the magnification map from the lens model of \citet{Diego16}. The white box encloses counter-images A and B of the $z$=5.75 LAE A370-L57 (green circles), while the white and pink circles mark the 5'' error circles around the expected position for counter-image C according to the lens models of \citet{Diego16} and \citet{Lagattuta17}, respectively. Right: enlarged view of the area inside the white box as seen in the RGB composite image (top), SHARDS F823W17 direct image (middle), and F823W17 residual frame after subtraction of a PSF-matched version of the F814W image (bottom). The radius of the green circles is 1".\label{fig:magnimap}}
\end{figure*}

\begin{figure*} 
\includegraphics[width=18cm,trim={0.5cm 0 0.5cm 0}]{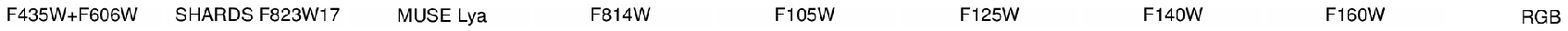}
\includegraphics[width=18cm,trim={0.5cm 0 0.5cm 0}]{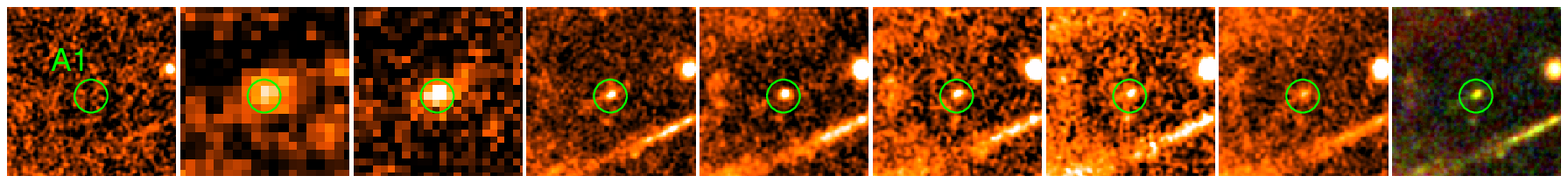}
\includegraphics[width=18cm,trim={0.5cm 0 0.5cm 0}]{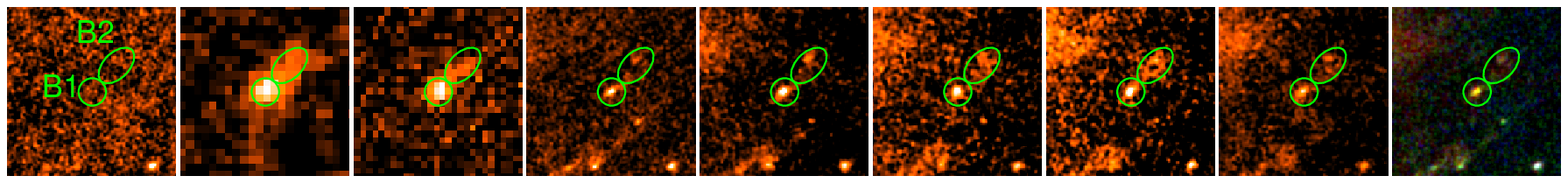}
\caption[]{4.5''$\times$4.5'' cutouts centered at the coordinates of counter-images A (top row) and B (bottom row). The pixel scale is 0.03'' for all HST bands, 0.25'' for the SHARDS F823W17 image, and 0.20'' for the MUSE Ly$\alpha$ image. The HST images have been smoothed with a Gaussian filter (radius = 3 pixels). The green ellipses indicate the apertures used for photometry.\label{fig:stamps}}
\end{figure*} 

We visually inspected the residual image to select emission line candidates. Among them we found a pair of elongated sources forming an incomplete arc, with their cores separated by $\sim$7'' and centered at (J2000.0) 2$^h$39$^m$51.67$^s$ -1$^o$35$^m$16.1$^s$ and 2$^h$39$^m$51.26$^s$  -1$^o$35$^m$12.6$^s$ (sources A and B, respectively, see Figure \ref{fig:magnimap}). 

The two sources are well detected ($\sim$10$\sigma$) in the SHARDS F823W17 filter (m$\sim$24.7), but are much fainter in F814W (m$\sim$27.3), implying the presence of an emission line with a large EW. In addition, the sources are detected in all the WFC3 bands but missing in a combination of the F435W and F606W images (Figure \ref{fig:stamps}), which suggests a strong break of the continuum emission at $\sim$8000 \AA{}, consistent with the emission line being Ly$\alpha$ at redshift $z$$\sim$5.7. 

The lens models by D16 and L17 confirmed that the two sources are counter-images of the same high-redshift galaxy. In the following we use the designation A370-L57 to refer to the galaxy itself and  A, B, for its counter-images.
D16 predicted a geometric redshift $z$$\sim$5.5 based on the relative positions of A and B. Subsequently, L17 confirmed the detection of Ly$\alpha$ in the MUSE spectrum and determined a spectroscopic redshift $z$=5.7505.

There are three MUSE pointings covering the counter-images A and B. In the synthetic F814W and F823W17 images extracted from the data cubes both counter-images are undetected due to the relatively shallow continuum sensitivity of MUSE observations, but they clearly show up in a synthetic narrow-band (8199--8215 \AA) image tailored to capture only the Ly$\alpha$ line (see Figure \ref{fig:stamps}). The seeing that we estimate in the MUSE Ly$\alpha$ images is slightly better compared to SHARDS F823W17 (FWHM$\sim$0.6''--0.7''). 

Both lens models predict high magnification at the observed position of the counter-images A ($\mu_{D16}$=10.7$\pm$1.9, $\mu_{L17}$=15.6$\pm$1.3) and B ($\mu_{D16}$=10.7$\pm$2.0, $\mu_{L17}$=16.6$\pm$1.2), with the critical line crossing between them (see top panel in Figure \ref{fig:magnimap}).
The stated uncertainties include only statistical errors, which for D16 are obtained from the dispersion of estimates for the set of 10 models tested, while L17 determines statistical errors using MCMC sampling with 5000 model realizations.
While the total magnification is $\sim$50\% larger in L17, the magnification ratios for the counter-images A and B are consistent between the two models ($\mu_A/\mu_B$ = 1.00$\pm$0.26 and 0.94$\pm$0.10 for D16 and L17, respectively). This is larger than the ratio 0.79$\pm$0.14 (median and dispersion) that we measure from the HST photometry in the five bands with detections (see \S\ref{sec:SED}), but consistent within the uncertainties.

The geometry of the lens determines that a third counter-image (C) should appear near the coordinates 2$^h$39$^m$57.18$^s$ -1$^d$34$^m$54.6$^s$ (D16 model) or 
2$^h$39$^m$56.80$^s$ -1$^d$34$^m$52.5$^s$ (L17 model). The two positions are $\sim$6'' apart. 
The relatively large offset between the two predicted positions (D16 and L17) of image C can be understood as a combination of factors, with perhaps the most important, the fact that image C was not included as a constraint in the lens models. Also, the image is expected to fall at the edge of the cluster core region, where there are few existing constraints. Such large offsets between predicted and observed positions are often found along features in giant elongated arcs, where small angular distances in the source plane translate into large angular distances in the image plane.

The predicted magnification for image C is $\mu$$\sim$3--4. This implies that C is $\sim$1.5 magnitudes fainter than A and B (that is, $m$$\sim$26.2 in F823W17 and $m$$\sim$28.8 in F814W), which is close to the detection limit of both SHARDS-FF and HFF observations.
In a stack of the F814W and the four WFC3 filters we find several faint sources within a 5'' error circle of the coordinates predicted by either the D16 or L17 models (white and pink circles in Figure \ref{fig:magnimap}). However, none of them shows any significant flux in F823W17 and colors consistent with the spectral energy distribution (SED) of A and B.
Unfortunately the expected coordinates for C are outside the area covered by available MUSE observations.

\section{Observed properties}\label{sec:obsproperties}
\subsection{Morphology}\label{sec:morphology}

\begin{figure}\hspace{-0.3cm} 
\includegraphics[width=9.0cm]{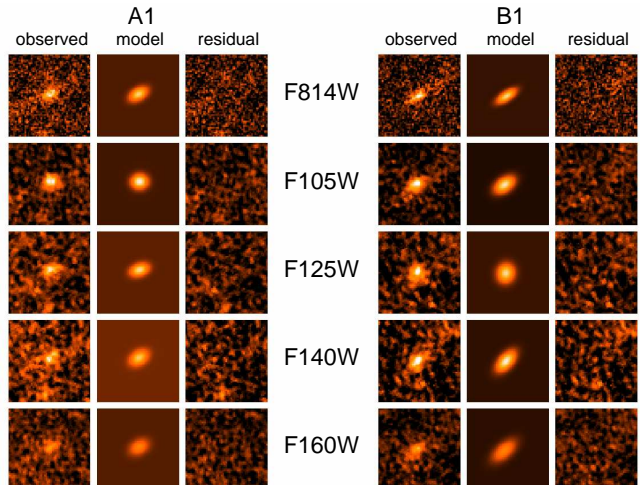}
\caption[]{Modeling of the observed profiles for sources A1 (left) and B1 (right) in the five HST bands with detections. The stamps are 1.5''$\times$1.5'' in size. The left column is the observed profile, the central column is the best-fitting 2D Gaussian plus constant background model, and the right column is the residual.\label{fig:psf2dfit}}
\end{figure}

\begin{figure}\hspace{-0.3cm} 
\includegraphics[width=9.0cm]{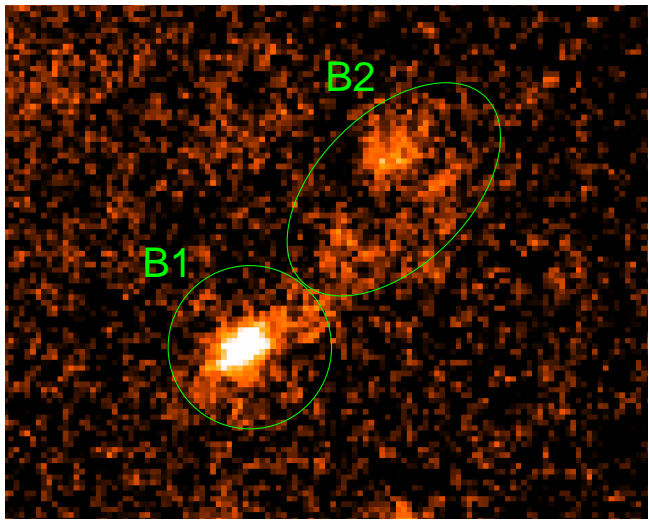}
\caption[]{Detection image showing the complex substructure in source B, consisting of an elongated core (B1) and an extended tail with multiple clumps (B2). The image combines individual images in the F814W and all four WFC3 filters with weights 0.5, 0.125, 0.125, 0.125, and 0.125. The radius of the B1 circle is 0.45''.\label{fig:detectionimage}}
\end{figure}

The apparent morphology of A370-L57 in the two detected counter-images (A and B) is dominated by the shear of the gravitational lens.
In the SHARDS F823W17 image, A and B are clearly elongated in the direction of shear with tails $\sim$1'' long extending in opposite directions. This extension is more evident in B, but this might be due to higher background near A from the outer regions of a nearby cluster member galaxy. 
The shape of the counter-images is consistent between the three MUSE Ly$\alpha$ images and SHARDS F823W17 after accounting for the variation in seeing. 

In the HST images, A and B are resolved into several components. Most of the flux arises from a very compact region (labelled A1 and B1 for counter-images A and B, respectively) that matches the peak of the emission in the SHARDS F823W17 and MUSE Ly$\alpha$ bands.

The profiles of A1 and B1 are elongated in the direction of shear by the lens due to a large shear factor (see Table \ref{table:magnification}). The elongation is most evident in the F814W band, probably because of the smaller PSF (0.09'' compared to 0.18--0.19'' in the WFC3 bands). 
We have measured the tangential and radial FWHM of A1 and B1 in all the HST bands by fitting elliptical 2D Gaussians on 1.5''$\times$1.5'' stamps. The results are shown in Table \ref{table:sizes} and Figure \ref{fig:psf2dfit}. 
Because the observed FWHM varies significantly among the WFC3 filters, we take their average as representative for the FWHM of the stellar emission, and its standard deviation as the uncertainty.
In \S\ref{sec:size} we estimate the physical size of A370-L57 from these FWHM measurements after correcting for PSF and the magnification by the lens.

In addition to this compact core, other fainter components likely contribute to the Ly$\alpha$ emission tail detected in the SHARDS F823W17 and MUSE data. 
These are most evident in counter-image B, probably due to the lower background.
In a detection image that combines the F814W band and all four WFC3 bands (Figure \ref{fig:detectionimage}), we recognize significant substructure with multiple clumps $\sim$1--1.5'' NW of B1. Since most of the clumps are undetected individually in single-filter HST images, and their Ly$\alpha$ emission cannot be isolated in the SHARDS F823W17 or MUSE Ly$\alpha$ images, we have considered them as a single source (B2) in the following. 
Because B2 is situated approximately along the major axis of B1, and at least some of its clumps are at the same redshift (see \S\ref{sec:lya}), it is likely that B2 is a companion of A370-L57 or even a distinct star forming region in the same galaxy. The projected distance of 1.3'' between the centers of B1 and B2 translates into $\sim$1.1 and $\sim$0.6 kpc for the D16 and L17 models, respectively.

\subsection{Spectral energy distribution}\label{sec:SED}

\begin{figure} 
\includegraphics[width=8.4cm]{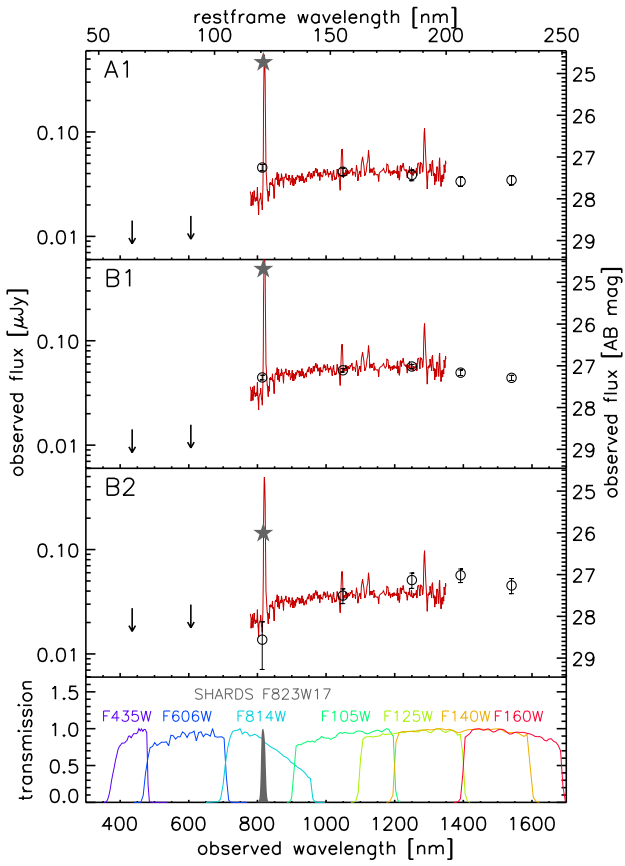}
\caption[]{Broadband spectral energy distributions for the regions enclosed by apertures A1, B1, and B2. The downward pointing arrows represent 3-$\sigma$ upper limits in the F435W and F606W bands. The grey star indicates the flux density in the SHARDS F823W17 filter. The red solid line, shown as a visual guide, is the composite spectrum of ten UV-selected sub-L* galaxies at $z$$\sim$3 from \citet{Amorin17}, scaled to match the F105W flux density of each source.
The bottom panel shows the (normalized) transmission curves of all the filters.\label{fig:arc-seds}}
\end{figure} 

We obtain photometry for A1, B1, and B2 using the apertures shown as green ellipses in Figure \ref{fig:stamps}. The same apertures are used for all filters. Aperture corrections are calculated for each filter using an empirical PSF derived from the combined images of 3 to 4 stars in the field.
Since B1 and B2 are blended at the resolution of SHARDS F823W17, we compute the contamination in B2 from the PSF wings of B1 by rotating the B2 aperture with center in B1 by 90, 180, and 270 degrees, and taking the mean of the three fluxes.
The resulting photometry is shown in Table \ref{table:photometry}. 

Figure \ref{fig:arc-seds} shows the observed SEDs for A1, B1, and B2. The flux ratio A1/B1 is $\sim$1 for the F814W and F823W17 bands, but between 0.7 and 0.8 in the WFC3 bands. The difference is too large to be explained by photometric errors alone. This is striking since gravitational lensing is achromatic. 
A possible interpretation could be different physical sizes of the Ly$\alpha$ and UV continuum emitting regions, or an offset between them. This is because the flux in the F814W and F823W17 filters is dominated by the Ly$\alpha$ line (see \S\ref{sec:lyaew}), while the WFC3 bands trace the UV continuum. Different morphologies for the two spectral components could translate into different values for the effective magnification if substructure in the lens causes a large local magnification gradient at the position of one of the images.
In our particular case, the presence of cluster member galaxies close to image A makes it difficult to obtain an accurate estimate of magnification in that region. However, the dispersion in the magnification estimates from the lens models should be indicative of the uncertainty in the magnification. Given these uncertainties, we find the observed flux ratios to be consistent with the model-predicted value in all the bands.
 
We estimate the spectral index $\beta$ of the rest-frame UV continuum (F($\lambda$) $\propto$ $\lambda^\beta$) from the observed fluxes in the F105W and F160W filters. Their effective rest-frame wavelengths are $\sim$1550 \AA{} and $\sim$2370 \AA{}, respectively, at $z$=5.75.
We obtain $\beta$ = -2.4$\pm$0.2, -2.4$\pm$0.1, and -2.3$\pm$0.2 for A1, B1, and B2, respectively.

\subsection{Lyman $\alpha$ emission}\label{sec:lya}

\begin{figure} 
\includegraphics[width=8.6cm]{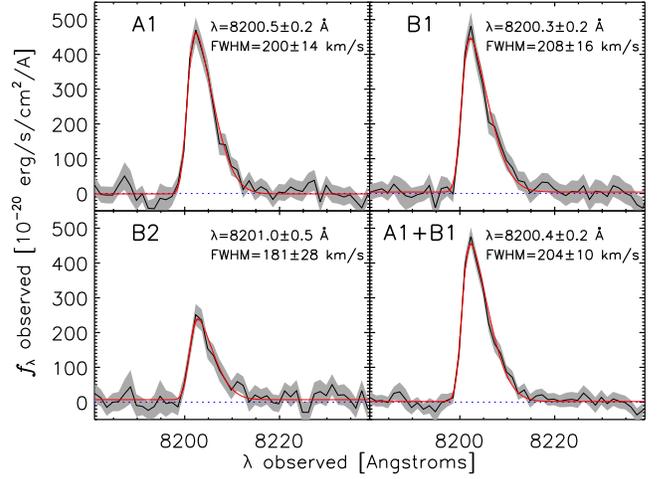}
\caption[]{Ly$\alpha$ profiles for sources A1, B1, B2, and the combined spectrum of A1 and B1. Each spectrum is the combination of the three individual spectra extracted on different MUSE data cubes. The shaded area represents the 1-$\sigma$ uncertainties, while the red solid line is the best-fitting model consisting of a half-Gaussian convolved with the instrumental profile. The derived peak wavelength and FWHM corrected for instrumental broadening are shown in the top right corner.\label{fig:MUSE-spectra}}
\end{figure}

We extract 1-D spectra from the individual MUSE data cubes for A1, B1, and B2, taking the same apertures used for photometry in \S\ref{sec:SED}.
We remove the residual background by subtracting the median spectrum in an annulus with inner and outer radii of 1.5'' and 2.5'', respectively.
For source B2, we compute and subtract the contamination from the PSF wings of B1 as in \S\ref{sec:SED}. Then we combine for each aperture the spectra from the three data cubes using a weighted average. 

The only spectral feature other than the residuals from telluric lines is the Ly$\alpha$ line, which peaks at 8202.3 \AA. 
There is no detection of the continuum at either side of Ly$\alpha$, which is consistent with the flux density estimated from the broadband images ($f_\lambda$ $\sim$0.5--1$\times$10$^{-20}$ erg cm$^{-2}$ s$^{-1}$ \AA$^{-1}$ redwards of Ly$\alpha$).
The line is detected at a $\sigma$$\sim$20 level in A1 and B1 and $\sigma$$\sim$6 in B2. 
The profile of the line is asymmetric with a broader wing on the red side (see Figure \ref{fig:MUSE-spectra}). 
A model consisting of a half Gaussian convolved with the instrumental profile ($FWHM_{instr}$=92.3 km s$^{-1}$) accurately reproduces the observed profile of the line in all three sources. The FWHM of the line (corrected for instrumental broadening) is 200$\pm$14 km s$^{-1}$ , 208$\pm$16 km s$^{-1}$, and 181$\pm$28 km s$^{-1}$ for A1, B1, and B2, respectively.
These line widths are small compared to brighter LAEs at this redshift, but similar to the widths measured by \citet{Karman17} for faint LAEs at 3$<z<$6. This is consistent with a continuation of the correlation found at higher luminosities between the width and luminosity of Ly$\alpha$ \citep{Hu10,Henry12}.

The peak of the half Gaussian in the best fitting model for the combined spectra of A1 and B1 is at $\lambda_c$=8200.4$\pm$0.2, implying a redshift of $z$=5.746. This probably overestimates slightly the actual redshift of A370-L57 since the peak of Ly$\alpha$ is usually redshifted relative to the systemic velocity of the galaxy. 
The profile for B2 is further redshifted by just 0.6$\pm$0.5 \AA, therefore consistent with being at the same redshift. 
 
We compute the Ly$\alpha$ flux by direct integration of the spectrum in the range 8199--8215 \AA. We obtain 3.28$\pm$0.16, 3.21$\pm$0.17, and 1.29$\pm$0.21 $\times$10$^{-17}$ erg cm$^{-2}$ s$^{-1}$ for A1, B1, and B2 respectively. The stated errors include only the photometric uncertainty. The uncertainty in the absolute flux calibration and the aperture correction introduce an additional systematic error of $\sim$20\% for luminosity-dependent properties.

\subsubsection{Ly$\alpha$ equivalent width}\label{sec:lyaew}

Since the continuum is undetected in the MUSE spectrum of A370-L57 on both sides of Ly$\alpha$, we measure the Ly$\alpha$ EW using the photometry in the SHARDS F823W17 and ACS F814W filters.
For this we model the spectrum of the source as the linear combination of two components, namely the Ly$\alpha$ line and the UV continuum:
\begin{equation}
f_\nu(\lambda) = a F_{Ly\alpha}(\lambda) + b F_{cont}(\lambda) 
\end{equation} 
\noindent where $F_{Ly\alpha}(\lambda)$ is a half-Gaussian peaking at 1216 \AA{} rest-frame with HWHM = 100 km s$^{-1}$, and $F_{cont}(\lambda)$ is a power-law with the spectral index $\beta$=-2.4 that we measured in \S\ref{sec:SED}. 
At $z$=5.75, the stellar continuum bluewards of Ly$\alpha$ is depleted by $\sim$90\% on average due to Gunn-Peterson absorption in a dense Ly$\alpha$ forest caused by intervening HI clouds \citep{Madau95,Fan06,Meiksin06}. We simulate this absorption by decreasing the flux in the continuum of our model by a factor 10 at rest-frame wavelengths between 912 and 1216 \AA. We also set the flux of the model spectrum at zero for wavelengths shorter than the Lyman limit (912 \AA).

To obtain the values of the free parameters $a$ and $b$ we convolve each of the components (redshifted to the observed frame) with the transmission profiles of the two filters, and solve the resulting 2$\times$2 linear system. 
We account for the photometric errors in the F814W and F823W17 flux densities by performing Monte Carlo simulations in which they are varied within their uncertainties. 
The flux in Ly$\alpha$ determined with this method is 3.4$\pm$0.2, 3.6$\pm$0.2, and 1.1$\pm$0.2 $\times$10$^{-17}$ erg s$^{-1}$ cm$^{-2}$ for A1, B1, and B2, respectively. The contribution from Ly$\alpha$ to the total flux density is $\sim$75\% and $\sim$98\% in the F814W and F823W17 filters, respectively.
The Ly$\alpha$ fluxes obtained with this method are in good agreement with the values from the MUSE spectra considering the 20\% systematic uncertainty.
We obtain rest-frame Ly$\alpha$ EWs of 360$^{+120}_{-80}$, 480$^{+140}_{-90}$, and 280$^{+360}_{-130}$ \AA{ } for A1, B1, and B2. A weighted average of A1 and B1 gives a maximum likelihood estimate for A370-L57 of EW$_0$(Ly$\alpha$)=420$^{+180}_{-120}$ \AA.

Our EW estimates take into account the attenuation of the continuum bluewards of Ly$\alpha$ due to IGM absorption, but not the IGM absorption on the Ly$\alpha$ line, which is also affected as manifested by the stronger asymmetry between the blue and red wings in high redshift galaxies compared to low redshift ones \citep{Hu04,Shimasaku06}.
Accounting for the IGM absorption on Ly$\alpha$ is complicated, since it is strongly dependent on the intrinsic velocity profile of the line and the clumpiness of the IGM in the vicinity of the galaxy. 
Even without this correction, the observed value in A370-L57 is higher than the maximum theoretical EW$_0$(Ly$\alpha$)=240 \AA{} for a "normal" population with a \citet{Salpeter55} IMF \citep{Charlot93,Malhotra02}, and it is at the extreme high end of the EW distribution at its redshift. As a comparison, only $\sim$25\% and $\sim$5\% of narrow-band selected LAEs at $z$=5.7 have apparent EW(Ly$\alpha$)$>$ 130 and 300 \AA, respectively \citep{Shimasaku06,Ouchi08}. 

Stellar population synthesis models can produce such high Ly$\alpha$ EW by assuming a top heavy IMF, very low metallicity and/or a very young burst of star formation $<$10$^7$ yr \citep[e.g.][]{Schaerer03}. In \S\ref{sec:sedfitting} we compare stellar population models with the observed SED.

\section{Magnification-corrected physical properties}\label{sec:delensed}

Table \ref{table:magnification} shows the estimated values for the radial ($\mu_{r}$), tangential ($\mu_{t}$), and total magnification ($\mu$ = $\mu_{t}\mu_{r}$), and for the shear factor ($S$ = $\mu_{t}/\mu_{r}$) for A1, B1, and B2.
The stated uncertainties include only statistical errors. Systematic uncertainties may be significantly larger, as evidenced by the total magnifications being $\sim$50\% larger in L17 compared to D16.

Near the critical curves, discrepancies in the magnification of up to a factor $\approx 2$ between different lens models are typical even in cases where the predicted critical curves are perfectly consistent between models. These discrepancies are related to  differences in the inferred location of the background sources that can differ from model to model. L17 uses more spectroscopic redshifts so one would expect D16 to be more affected by the uncertainty on the redshift of the background sources than L17.

Since we have no means to test which of the two parameter sets is closer to the actual values, all magnification-dependent quantities are subject to this uncertainty. 
For clarity, we have not propagated this uncertainty into error estimates for other properties. 

In the following all magnification-corrected quantities refer to the L17 model. To obtain the corresponding values in the D16 model it suffices to multiply by 1.5 for the luminosity, SFR, and stellar mass, by 2.0 for the (tangential) half-light radius, and by 0.82 for the surface SFR density. A summary of the main physical properties of the galaxy can be found in Table \ref{table:physprop}.

\subsection{UV continuum and Ly$\alpha$ luminosity}\label{sec:delensedLum}

We estimate the absolute magnitude at rest-frame 1600 \AA{}, M$_{\rm{UV}}$, from the flux density in the F105W filter. Given total magnifications (L17) $\mu$ = 16.5, 15.4, and 12.1, we get M$_{\rm{UV}}$ = -16.27$\pm$0.06, -16.59$\pm$0.03, and -16.00$\pm$0.05 for A1, B1, and B2, respectively. 

We take M$_{UV}$$\sim$-16.5 as a compromise value between A1 and B1 for the absolute magnitude of A370-L57. This implies that A370-L57 and B2 are two orders of magnitude fainter than the characteristic UV luminosity at this redshift \citep[M$^*_{\rm{UV}}$=-20.95;][]{Bouwens15b}. 
The Ly$\alpha$ flux measured on the MUSE spectrum translates into a magnification-corrected luminosity of 7.7$\pm$0.4$\times$10$^{41}$ erg s$^{-1}$ and 2.6$\pm$0.4$\times$10$^{41}$ erg s$^{-1}$ for A370-L57 and B2, respectively.

Over the past few years, several other LAEs have been found at $z$$\sim$5--6 with comparably low Ly$\alpha$ and UV continuum luminosities. Like our system, all of these objects have also been magnified by a massive galaxy cluster. \citet{Karman17} presented a detailed study of a sample of lensed low-luminosity LAEs in the HFF cluster AS0163 detected through a blind search with MUSE. \citet{Caminha17} reported the spectroscopic confirmation of 22 sources within the redshift range $z$=3.2--6.1 in the HFF cluster MACS 0416, most of them low luminosity LAEs. The least luminous one ($z$=6.1, M$_{UV}$=-15) was recently discussed by \citet{Vanzella17}.

\subsection{Star formation rate}

We can estimate the SFR from the magnification-corrected UV luminosity using e.g. the conversion given by \citet{Kennicutt98}:
\begin{equation}
SFR_{UV}{ }[M_\odot{ }yr^{-1}] = 7.7\times10^{-29} L_\nu(1600\AA) [erg{ }s^{-1}{ } Hz^{-1}] 
\end{equation} 
\noindent where the assumptions are solar metallicity, a constant SFR in the last 100 Myr, and a \citet{Salpeter55} IMF, which we have converted to a \citet{Chabrier03} IMF by applying a factor 0.55 \citep[see e.g.][]{Erb06}. 
This implies SFR$_{UV}$ $\sim$0.13 and $\sim$0.08 M$_\odot$ yr$^{-1}$ for A370-L57 and B2, respectively. 

While the Ly$\alpha$ luminosity is a poor indicator of the SFR in galaxies due to radiative transfer effects, it is interesting for comparison, since in most LAEs it is the only SFR indicator available. Following \citet{Kennicutt98}, the rate of production of ionizing photons N(LyC) and the SFR are related by:
\begin{equation}\label{eq:SFRNLyC}
SFR{ }[M_\odot yr^{-1}] = 5.9\times10^{-54} N(LyC){ }[s^{-1}]
\end{equation}

\noindent where we have again applied a factor 0.55 for conversion to a \citet{Chabrier03} IMF. Assuming case B recombination and a gas temperature $T$$\sim$10$^4$ K, the intrinsic Ly$\alpha$ luminosity and N(LyC) are related by \citep[e.g.][]{Dijkstra14}: 
\begin{equation}\label{eq:LyaNLyC}
L(Ly\alpha)_{int} = 0.68{ } h\nu{ }(1-f_{esc}^{LyC}){ }N(LyC) 
\end{equation}

\noindent where $h\nu$ is the energy of an individual Ly$\alpha$ photon and $f_{esc}^{LyC}$ is the fraction of LyC photons that escape to the IGM. $f_{esc}^{LyC}$ is hard to estimate at $z$$>$4 because all LyC emission is absorbed in the IGM. Extrapolation from low redshift analogs and indirect estimates suggest that $f_{esc}^{LyC}$$\sim$0.1--0.2 is a reasonable range for LAEs at $z$$\gtrsim$6 \citep{Faisst16}. Assuming $f_{esc}^{LyC}$=0.15 and no extinction, from Eq. \ref{eq:SFRNLyC} and \ref{eq:LyaNLyC} we obtain:

\begin{equation}
SFR_{Ly\alpha} [M_\odot yr^{-1}] = 6.3\times10^{-43} L(Ly\alpha) [erg s^{-1}]
\end{equation}

\noindent The resulting SFR$_{Ly\alpha}$ are $\sim$0.48 and $\sim$0.16 M$_\odot$ yr$^{-1}$ for A370-L57 and B2. This is a factor 3.7 and 2 higher, respectively, than obtained from the UV continuum, which is remarkable since the observed Ly$\alpha$ luminosity probably underestimates the intrinsic value due to absorption by dust grains in the ISM of the galaxy.

This suggests that the stellar population is significantly younger and less metallic than the Kennicutt relation assumes \citep[see e.g.][]{Verhamme08}.
In \S\ref{sec:sedfitting} we obtain a more realistic estimate of the SFR using the best-fitting stellar population model. 

\subsection{Size of the Ly$\alpha$ and continuum emitting regions}\label{sec:size}

Since B2 is extremely faint and consists of several clumps barely detected individually, we obtain an estimate of the galaxy size, parametrized by the half-light radius ($r_e$), only for A370-L57.

We corrected for PSF the FWHM measurements for A1 and B1 in Table \ref{table:sizes} by subtracting in quadrature the PSF FWHM. The values used are 0.093'' for F814W and 0.18--0.19'' for the WFC3 filters. 

The PSF-corrected sizes for F814W and WFC3 are comparable, at $\sim$0.25--0.30'' and $\sim$0.10'' in the tangential and radial directions, respectively. 
Due to the strong shear by the lens, this suggests that the galaxy is actually elongated in the radial (NE-SW) direction, with $r_e$$\sim$0.23--0.25 kpc compared to just $\sim$0.05--0.06 kpc in the tangential (NW-SE) direction.

To put these dimensions in perspective, we have computed the expected $r_e$ of our galaxy using the size-redshift-luminosity relation found by \citet{Shibuya15} in a sample of $\sim$190,000 galaxies at $z$=0--10:

\begin{equation}
r_e = B_z(1+z)^{\beta_z}\big{(}\frac{L_{UV}}{L_0}\big{)}^\alpha
\end{equation}
\noindent where $B_z$ = 6.9 kpc, $\beta_z$ = -1.20$\pm$0.04, $\alpha$ = 0.27$\pm$0.01, and $L_0$ is the luminosity corresponding to M$_{UV}$=-21.

For a magnification-corrected luminosity M$_{UV}$ = -16.5 we get $r_e$ = 0.25 kpc, which for a roughly circular galaxy translates into a half-light area of 0.20 kpc$^2$, compared to 0.042--0.051 kpc$^2$ in A370-L57. 
This implies that the surface brightness in A370-L57 is 4--5 times higher than predicted from the extrapolation of the \citet{Shibuya15} relation. 
However, recent results by \citet{Bouwens17} in a sample of highly magnified galaxies from four HFF clusters, favor a much steeper size-luminosity relation for ultra-faint galaxies at $z$=6--8, with $\alpha$=0.5 for (magnification corrected) m$>$29. At M$_{UV}$=-16.5, their relation implies $r_e$ = 0.02'' or 0.12 kpc at $z$=5.75. The predicted half-light area of 0.045 kpc$^2$ is within our confidence interval.

We note, however, that the radial FWHM measurement is well constrained only for the ACS/F814W image, while for the individual WFC3 bands it is compatible with an unresolved source at the 2-$\sigma$ level. It is conceivable that the Ly$\alpha$ emission is more extended in the NE-SW direction than the stellar emission. 
If we dismiss the radial FWHM measurement from WFC3 and assume that the actual shape of A370-L57 is roughly circular, then the tangential size would translate into a half-light area of just $\sim$0.01 kpc$^2$, or $\sim$5 times smaller than expected from the \citet{Bouwens17} relation. 

\section{Stellar population modeling}\label{sec:sedfitting}

\begin{figure} 
\includegraphics[width=8.6cm]{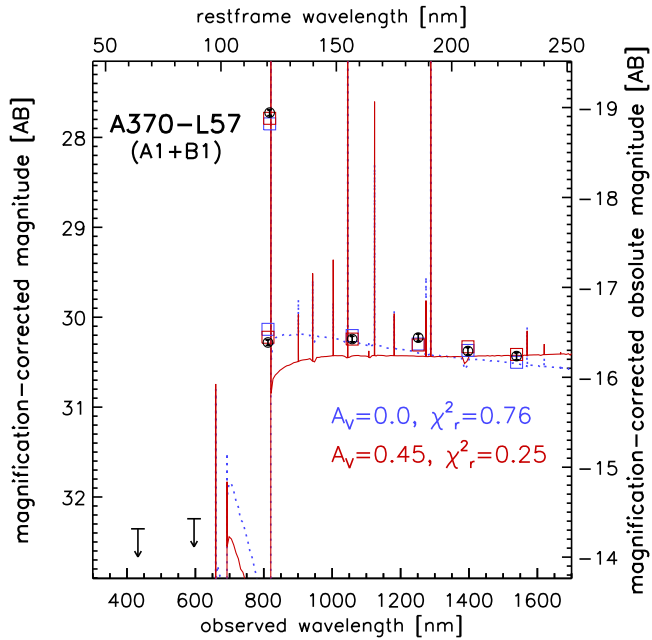}
\caption[]{Best-fitting stellar population models for A370-L57. The circles with error bars represent the observed photometry, corrected for magnification assuming the magnification values from the L17 model. Downward pointing arrows are 3-$\sigma$ upper limits in the F435W and F606W bands. The solid and dotted lines represent, respectively, the best-fitting model with and without extinction. The squared symbols represent the synthetic photometry obtained from convolution of the best-fitting model with the filter transmission curves.\label{fig:modelfits}}
\end{figure}

We have fitted the HST and SHARDS photometry of A370-L57 with stellar population synthesis models. 
For this purpose we use the Code Investigating GALaxy Emission \citep[CIGALE;][]{Noll09,Serra11}, in its \textit{python} implementation, \textit{pcigale}\footnote{//http://cigale.lam.fr} version 0.11.0 (see also \citet{Boquien16} for an updated description).

The stellar emission spectra are built from the \citet{Bruzual03} stellar library. We assume a \citet{Chabrier03} IMF and a delayed exponential star formation history, SFR($t$) $\propto$ $t e^{-t/\tau}$. We generate a grid of models covering a range of metallicities ($Z$=0.0001--0.02), ages ($t$/Myr=1--1000), and decay times ($\tau$/Myr=1--1000). 
From the number of Lyman continuum photons N(LyC) predicted by these models, \textit{pcigale} computes the nebular emission (lines and continuum) using dedicated CLOUDY \citep{Ferland98} templates based on the models of \citet{Inoue11}. The ionization parameter $\log U$ can take any value between -4.0 and -1.0 in steps of 0.5, and the escape fraction of LyC photos ranges between 0 and 0.5 in steps of 0.02. For a given value of $\log U$, the nebular emission is directly proportional to (1-$f_{esc}^{LyC}$)N(LyC).  We assume that the absorption of LyC photons by dust is negligible. 

To model the effect of the extinction on the stellar and nebular continuum we try two approaches. The first one assumes no extinction at all. The second one assumes a \citet{Calzetti01} extinction law with $R_V$ = 4.05 and free E(B-V) ranging between 0 and 0.25 magnitudes, and a foreground screen for the dust geometry.

The case with adjustable extinction obtains the best fit for a model with E(B-V)=0.11 mag ($A_V$=0.45 mag). Figure \ref{fig:modelfits} compares this model with the best-fitting one for E(B-V)=0 and the observed photometry.
We note that the continuum (stellar+nebular) for the model with extinction is significantly redder, as expected. However, in the synthetic photometry (model convolved with filter profiles) this is compensated to a large extent by the stronger nebular lines resulting from a younger age ($t$=2 Myr versus $t$=4 Myr for E(B-V)=0) and much higher SFR $\sim$4.6 M$_\odot$ yr$^{-1}$ for E(B-V)=0.11 and $\sim$0.26 M$_\odot$ yr$^{-1}$ for E(B-V)=0.0, see Table \ref{table:sedfit} for all the model parameters). 

While the reduced $\chi^2$, $\chi^2_r$ = $\chi^2$/(N$_{bands}$-1), is significantly better for the case with extinction (0.25 versus 0.76 for E(B-V)=0), both are consistent with the observations within their uncertainties. 
This highlights the degeneracy that affects the model parameters, even for such a simple model with a single stellar population. To mitigate this and to obtain realistic uncertainties for the model parameters, \textit{pcigale} performs a bayesian analysis that obtains marginalized posterior probability distribution functions (PDFs) for the parameters and derived quantities like the SFR. The expectation values and 1-$\sigma$ uncertainties are listed in the right column of Table \ref{table:sedfit}. 

The metallicity is well constrained and agrees within the uncertainties for the models with and without extinction. In both cases the best fit is obtained for $Z$=0.0004. The bayesian analysis indicates that the metallicity could also be higher, up to $Z$$\sim$0.004, while there is no lower limit (the lowest metallicity probed by the models, $Z$=0.0001, also gives excellent fits).
The best fits with and without extinction are obtained for $t$ = 2 and 4 Myr, respectively, but the PDFs indicate that the observations are compatible with somewhat older ages, between 2.5 and $\sim$10 Myr. 

It is interesting that in both cases, the best fit is obtained with the lowest values of $\tau$ and $f_{esc}^{LyC}$ available in the grid of models, and that $\log U$ takes the highest value in the case with extinction. Since further extending the parameter space would lead to unphysical values, this tensions suggests that some of the model assumptions might not be accurate. 
In particular, our assumption of a single burst of star formation is in all likelihood an oversimplification of the actual SFH. 
Any old ($\gtrsim$100 Myr) population in A370-L57 is largely unconstrained due to lack of meaningful upper limits redwards of $\sim$2500 \AA. Such population could contribute a significant fraction of the UV stellar continuum, but its nebular emission would be negligible. As a consequence, the net effect of an unaccounted for old(-ish) population is to redden the continuum and to decrease the EW of nebular lines. Accordingly, to reproduce the observed photometry the young population would shift to even lower values for $t$ and $Z$.

The lack of constraints for the old population makes the total stellar mass of the galaxy highly uncertain. Assuming that the young population dominates the mass budget and no extinction, we get a lower limit of $\sim$1.4$\times$10$^6$ M$_\odot$. A realistic upper limit for the mass of the young population would be the value obtained with free extinction, 4.5$\times$10$^6$ M$_\odot$. Therefore we take 3.0$\pm$1.5$\times$10$^6$ M$_\odot$, as our best estimate of the stellar mass of A370-L57. 

The instantaneous SFR for the best-fitting models differ by more than an order of magnitude between the cases with and without extinction, but the expectation values from the bayesian analysis agree within a factor $\sim$2, consistent with their uncertainties. We take as the final value the geometric mean of the two estimates, $\log (SFR/M_\odot yr^{-1})$ = -0.1$\pm$0.3. This is higher than obtained from the Ly$\alpha$ luminosity using the \citet{Kennicutt98} relation, but consistent within the uncertainties.

The corresponding specific SFR is $\log (sSFR/Gyr^{-1})$ = 2.4$\pm$0.4 (assuming that an older population does not contribute significantly to the total stellar mass) which corresponds to a stellar mass doubling timescale of $\sim$3 Myr. This specific SFR is high compared to normal star-forming galaxies at lower redshift, but consistent with the values found by \citet{Karman17} in $z$=5--6 LAEs with similar stellar masses.
The SFR density for A370-L57 ranges between $\sim$7 and $\sim$35 M$_\odot$ yr$^{-1}$ kpc$^{-2}$ depending on the actual elongation in the radial direction. 
This is in agreement with the SFR densities measured in a sample of  comparably low mass LAEs at $z$$\sim$3 by \citet{Amorin17}, and with the extremely small sizes inferred for strongly lensed $z$=2--8 sources by \citet{Bouwens17}.

\section{Origin of the Ly$\alpha$ emission in A370-L57}

Since the main observable pushing for a very young and low metallicity stellar population is the high EW of Ly$\alpha$, it is important to consider if a non-stellar source could contribute to the observed Ly$\alpha$ emission. 

Deep rest-frame UV spectroscopy of other LAEs with steep UV continua has revealed several sources with strong CIV and OIII] emission lines, but not HeII \citep[e.g.:][]{Stark15,Berg16,Mainali17}. This is consistent with the ionizing spectrum of the extremely hot stars expected in young and very metal poor populations, but not with the typical power-law spectrum of AGN \citep[see e.g.][]{Feltre16}.

While all the main UV lines are outside the wavelength range of the MUSE observations, we obtain an upper limit for the flux in the N{\sc v} 1239 \AA{} line of $<$1.5$\times$10$^{-18}$ erg cm$^{-2}$ s$^{-1}$ for A370-L57. This implies Ly$\alpha$/N{\sc v}$>$20, which together with the narrow Ly$\alpha$ profile and the fact that the source is resolved in both Ly$\alpha$ and the continuum, makes a significant contribution from a low luminosity AGN unlikely.

\begin{figure*} 
\includegraphics[width=6.1cm]{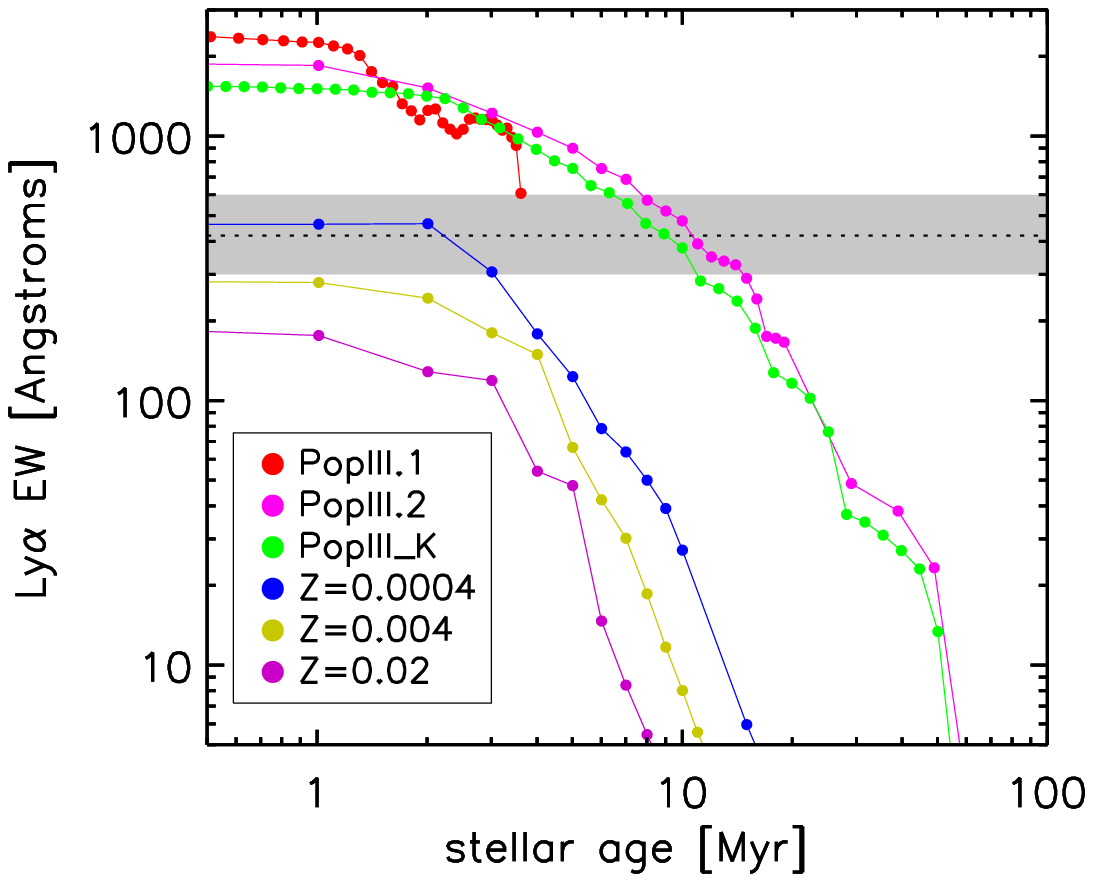}
\includegraphics[width=6.1cm]{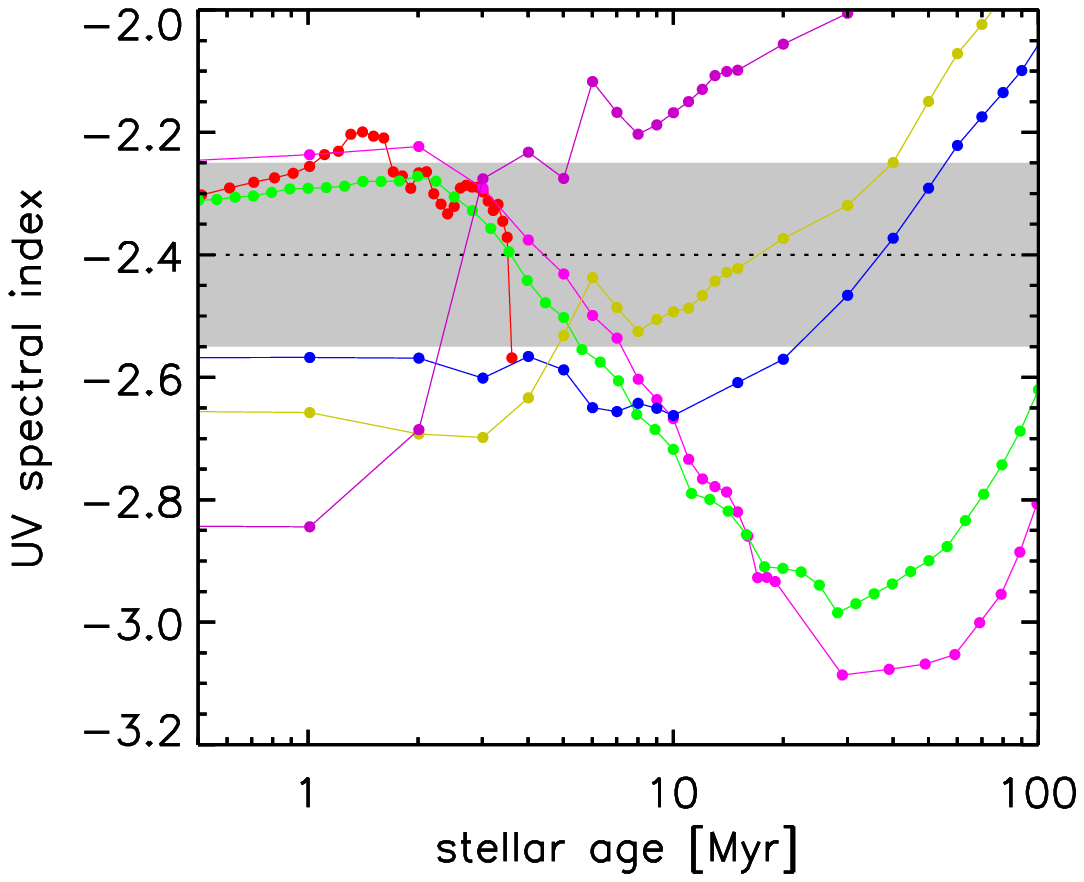}
\includegraphics[width=6.1cm]{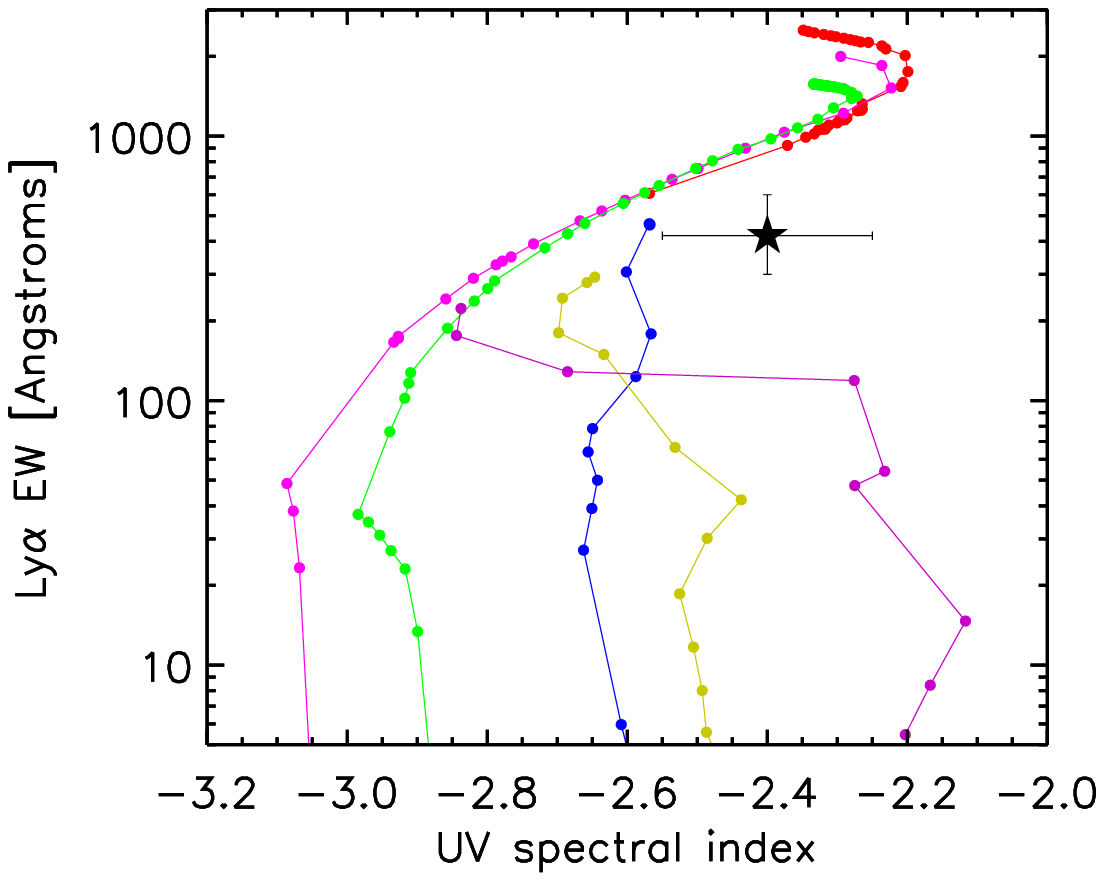}
\caption[]{Time evolution of the equivalent width of the Ly$\alpha$ line and the UV spectral index of a instantaneous burst for a range of metallicities. The tracks were generated using stellar population models obtained with Yggdrasil \citep{Zackrisson11}. The models use the \citet{Schaerer02} stellar library for PopIII stars and Starburst99 \citep{Leitherer99} for PopII and PopI stars. Metallicities range from zero (PopIII models) to solar. Extinction is assumed to be negligible, and the covering factor of the gas (which determines the strength of the nebular lines) is set at 100\%. The IMF is assumed to be that of \citet{Kroupa01} except for the PopIII.1 and PopIII.2 models, which assume extremely and moderately top-heavy IMFs, respectively \citep[see ][for details]{Zackrisson11}. The dotted lines in the left and center plots mark our best estimate for EW$_0$(Ly$\alpha$) and $\beta$ in A370-L57, and the gray areas represent the 1-$\sigma$ confidence interval. In the right panel, the same information is represented by the solid star with error bars.\label{fig:LyaEW-beta-models}}
\end{figure*}

Population III (PopIII) or extremely metal-poor ($Z$$<$10$^{-5}$) Population II (PopII) stars have been hypothesized to contribute at least a fraction of the Ly$\alpha$ emission of some LAEs with large Ly$\alpha$ EW found at $z$$\sim$6.5 \citep[e.g.][]{Kashikawa12,Sobral15,Rydberg17}.
PopIII stellar models predict that, for $t$$<$2 Myr, EW$_0$(Ly$\alpha$) can be as high as 3000 \AA{}\protect{ } \citep[][]{Schaerer03,Zackrisson11}, implying that only a modest contribution from a burst of PopIII stars would suffice to obtain the observed EW$_0$(Ly$\alpha$) even if the UV continuum and the stellar mass are dominated by normal PopI/PopII stars. 

The smoking gun for the presence of PopIII stars would be HeII 1640 \AA{} emission with an EW higher than $\sim$5 \AA \, \citep{Schaerer03}. One possible reason why no indisputable evidence of PopIII stars has been found so far is that spectroscopic searches of the elusive HeII 1640 \AA{} line have mostly targeted luminous LAEs due to observational limitations. Numerical simulations predict that PopIII stars form preferentially in low mass systems \citep[M$_*$/M$_\odot$$<$10$^{6.5}$;][]{Pallotini15}, and less massive galaxies are also more likely to maintain patches of pristine gas until later epochs. 

Because A370-L57 is among the least massive known galaxies at $z$$\sim$6 and it has extreme EW$_0$(Ly$\alpha$), it is instructive to check how its observed properties compare to both PopIII and normal PopI/II stellar models.

We have used the Yggdrasil models \citep{Zackrisson11} to illustrate how EW$_0$(Ly$\alpha$) and $\beta$ depend on the metallicity and age of the stellar population for both normal and PopIII populations (Figure \ref{fig:LyaEW-beta-models}). 
The left panel shows the evolution of EW$_0$(Ly$\alpha$) for a instantaneous burst with different metallicities. The tracks represent the theoretical maximum for each population, which implies A$_V$=0, $f_{esc}^{LyC}$=0, and no absorption of Ly$\alpha$ photons in the surrounding IGM.
From this figure it is immediately evident that only $Z$$\lesssim$4$\times$10$^{-4}$ reproduces the observed value of EW$_0$(Ly$\alpha$) in A370-L57, while $Z$=4$\times$10$^{-3}$ would also be consistent within the uncertainty, but only at extremely young age ($t$$\lesssim$2 Myr). Ages older than $\sim$10 Myr are ruled out for any metallicity (including PopIII).

The dependence of $\beta$ with the age and metallicity of the population (central panel) is more complex. At very young age ($t\lesssim$2 Myr), $\beta$ is almost independent on $t$ and higher at lower metallicity (due to increased nebular emission). However, as the population ages and the nebular emission decreases, $\beta$ increases in the $Z$$>$0 populations but decreases for PopIII. By $t$$\sim$10 Myr the nebular emission has become negligible and $\beta$ is monotonously higher for higher metallicity. 
The observed value for $\beta$ in A370-L57 is consistent with almost any metallicity at some age. However, since $t$$>$3 Myr is ruled out for $Z$$>$4$\times$10$^{-4}$ by EW$_0$(Ly$\alpha$), the observed $\beta$ requires $Z$$\lesssim$4$\times$10$^{-4}$. 
The right panel rules out a PopIII-dominated SED for A370-L57, because for the observed $\beta$ the required EW$_0$(Ly$\alpha$) would be $\sim$1000 \AA.  
Despite this, our single population delayed exponential model is likely a gross oversimplification of the actual SFH of the galaxy. While the currently available data do not allow us to test more complex SFHs, a wider range of stellar ages and metallicities may be present. If older or more metallic stars contribute significantly to the UV continuum emission, younger and/or less metallic ones are required to compensate for their lower production rate of LyC photons in order to reproduce the observed Ly$\alpha$ luminosity. 

A conclusive diagnostic of the source of the Ly$\alpha$ emission in A370-L57 requires the determination of UV emission line ratios.
Recently, \citet{Vanzella16} measured line ratios CIV/Ly$\alpha$=0.25, HeII/Ly$\alpha$=0.067, and OIII]/Ly$\alpha$=0.16 in the spectrum of a lensed $z$=3.11 LAE with comparably low luminosity (M$_{UV}$=-17.0). If A370-L57 has similar line ratios, the CIV 1550 \AA, HeII 1640 \AA, and OIII] 1660,1666 \AA{} lines, among others, would be detectable with deep NIR spectroscopy on current 10-m class telescopes.\\

\section{Conclusions}

In this work we have taken advantage of strong lensing by the Abell 370 galaxy cluster in combination with deep imaging from the Hubble Frontier Fields and SHARDS Frontier Fields programs, and MUSE spectroscopy, to perform a detailed analysis of a Ly$\alpha$ emitter. This source would be fainter than magnitude 30 in the continuum if it was not magnified by the gravitational lens. 

The source, A370-L57, is a $z$=5.75 galaxy with a low Ly$\alpha$ luminosity ($L$(Ly$\alpha$)$\sim$10$^{42}$ erg s$^{-1}$) that however implies an extreme Ly$\alpha$ EW of $\sim$420 \AA. This value of the EW is exceptional among known galaxies in this luminosity range and redshift. The even fainter source B2, with similar properties and confirmed spectroscopically to be at the same redshift, is just $\sim$1 kpc away in projection, and could be another star-forming region in the same galaxy or a close neighbor. 

The physical properties of A370-L57 are similar to those of galaxies with comparable UV luminosity irrespective of their redshift or Ly$\alpha$ EW. That is: very compact size ($r_e$$<$100pc), high specific SFR ($sSFR$$\sim$2.5$\times$10$^{-7}$ yr$^{-1}$), high SFR density ($\Sigma_{SFR}$$\sim$7--35 M$_\odot$ yr$^{-1}$ kpc$^{-2}$), and blue UV spectrum ($\beta$$\sim$-2.4). 
However, the very high Ly$\alpha$ EW seems to require an extremely young ($t$$<$10 Myr) and low metallicity ($Z$$\lesssim$4$\times$10$^{-3}$) stellar population. This result is robust independently on the amount of extinction. We also find no evidence of AGN activity that could contribute to the Ly$\alpha$ emission.

The physical properties of A370-L57 and its strong magnification make it an interesting candidate to search for the spectroscopic signature of low metallicity ($Z$$<$10$^{-4}$) star formation. 
Deep NIR spectroscopy with sufficient sensitivity to detect diagnostic UV lines is challenging but feasible with recent instrumentation. 
 
\acknowledgements

We thank the anonymous referee for thoughtful comments and suggestions that helped improve this work. A.H.-C. and P.G.P.-G. acknowledge funding by the Spanish Ministry of Economy and Competitiveness (MINECO) under grants AYA2012-31277, AYA2015-70815-ERC, and AYA2015-63650-P. 
J.M.D acknowledges support of the projects AYA2015-64508-P (MINECO/FEDER, UE), AYA2012-39475-C02-01 and the consolider project CSD2010-00064 funded by MINECO.
D.J.L. and J.R. acknowledge support from the ERC starting grant 336736-CALENDS.
R.A.M. acknowledges support by the Swiss National Science Foundation.
A.A.-H. acknowledges funding by the MINECO grant AYA2015-64346-C2-1-P which is partly funded by the FEDER program. HDS acknowledges funding from the program ANR-T-ERC ASTROBRAIN.
P.S. acknowledges funding from the Swiss National Science Foundation under grant P2GEP2\_165426. J.M.R.E. acknowledges funding by the MINECO grant AYA2015-70498-C2-1-R. H.D. acknowledges financial support from the Spanish 2014 Ram\'on y Cajal program MINECO RYC-2014-15686.

\clearpage

\begin{deluxetable}{c|cc|cccc|cccc} 
\centering
\tabletypesize{\small}
\tablewidth{0pc}
\tablecolumns{12}
\centering
\tablecaption{Magnification properties from lens models\label{table:magnification}}
\tablehead{\colhead{} & \colhead{} & \colhead{} & \multicolumn{4}{c}{\citet{Diego16}} & \multicolumn{4}{c}{\citet{Lagattuta17}} \\
\colhead{Source} & \colhead{RA (J2000)} & \colhead{DEC (J2000)} & \colhead{$\mu_r$\tablenotemark{a}} & \colhead{$\mu_t$\tablenotemark{b}} & \colhead{$\mu$\tablenotemark{c}} & \colhead{S\tablenotemark{d}} & \colhead{$\mu_r$\tablenotemark{a}} & \colhead{$\mu_t$\tablenotemark{b}} & \colhead{$\mu$\tablenotemark{c}} & \colhead{S\tablenotemark{d}}}
\startdata
A1 & 2$^h$39$^m$51.67$^s$ & -1$^o$35$^m$16.1$^s$ & 1.3$\pm$0.2 & -7.0$\pm$1.0 & 10.7$\pm$1.9 & -5.4$\pm$1.1 & 1.18$\pm$0.21 & -13.1$\pm$1.3 & 15.6$\pm$1.3 & -11.1$\pm$2.3 \\
B1 & 2$^h$39$^m$51.26$^s$ & -1$^o$35$^m$12.6$^s$ & 1.3$\pm$0.2 &  7.0$\pm$1.1 & 10.7$\pm$2.0 &  5.4$\pm$1.2 & 1.16$\pm$0.15 &  14.2$\pm$1.2 & 16.6$\pm$1.2 &  12.3$\pm$1.9 \\
B2 & 2$^h$39$^m$51.21$s$  & -1$^o$35$^m$11.9$^s$ &        &     &  8.4$\pm$1.0      &      &     &  & 12.1$\pm$1.1  &       \\ 
\enddata
\tablenotetext{a}{radial magnification}
\tablenotetext{b}{tangential magnification}
\tablenotetext{c}{total magnification $\mu$=$\mu_r$$\mu_t$}
\tablenotetext{d}{shear factor $S$=$\mu_t$/$\mu_r$}
\end{deluxetable}

\begin{deluxetable}{lc|r@{ $\pm$ }lr@{ $\pm$ }lr@{ $\pm$ }l|r@{ $\pm$ }lr@{ $\pm$ }lr@{ $\pm$ }l} 
\centering
\tabletypesize{\small}
\tablewidth{0pc}
\tablecolumns{10}
\centering
\tablecaption{Size measurements from 2D Gaussian fits\label{table:sizes}}
\tablehead{\colhead{} & \colhead{} & \multicolumn{6}{c}{A1} & \multicolumn{6}{c}{B1} \\
\colhead{Filter} & \colhead{PSF FWHM} & \multicolumn{2}{c}{major axis\tablenotemark{*}} & \multicolumn{2}{c}{minor axis\tablenotemark{*}} & \multicolumn{2}{c}{angle} & \multicolumn{2}{c}{major axis\tablenotemark{*}} & \multicolumn{2}{c}{minor axis\tablenotemark{*}} & \multicolumn{2}{c}{angle} \\
\colhead{} & \colhead{('')} & \multicolumn{2}{c}{('')} & \multicolumn{2}{c}{('')} & \multicolumn{2}{c}{(deg)} & \multicolumn{2}{c}{('')} & \multicolumn{2}{c}{('')} & \multicolumn{2}{c}{(deg)}}
\startdata
F814W & 0.093 & 0.29 & 0.02 & 0.18 & 0.01 & 149 &   4 & 0.32 & 0.02 & 0.14 & 0.01 & 147 &   2 \\
F105W & 0.181 & 0.24 & 0.01 & 0.21 & 0.01 & 175 &  19 & 0.33 & 0.02 & 0.20 & 0.01 & 145 &   4 \\
F125W & 0.185 & 0.28 & 0.03 & 0.18 & 0.02 & 160 &   9 & 0.28 & 0.02 & 0.24 & 0.02 & 107 &  26 \\
F140W & 0.187 & 0.32 & 0.04 & 0.22 & 0.03 & 152 &  12 & 0.35 & 0.03 & 0.19 & 0.02 & 132 &   5 \\
F160W & 0.190 & 0.31 & 0.03 & 0.22 & 0.02 & 150 &  12 & 0.42 & 0.03 & 0.22 & 0.02 & 139 &   4 \\
\enddata
\tablenotetext{*}{not corrected for PSF}
\end{deluxetable}

\begin{deluxetable}{lr@{ $\pm$ }lr@{ $\pm$ }lr@{ $\pm$ }l} 
\centering
\tabletypesize{\small}
\tablewidth{0pc}
\tablecolumns{10}
\centering
\tablecaption{Observed photometry\label{table:photometry}}
\tablehead{\colhead{} & \multicolumn{2}{c}{A1} & \multicolumn{2}{c}{B1} & \multicolumn{2}{c}{B2}}
\startdata
ACS F435W  &  \multicolumn{2}{c}{$>$28.96 (3$\sigma$)} & \multicolumn{2}{c}{$>$28.96 (3$\sigma$)} & \multicolumn{2}{c}{$>$28.96 (3$\sigma$)} \\
ACS F606W  &  \multicolumn{2}{c}{$>$28.85 (3$\sigma$)} & \multicolumn{2}{c}{$>$28.85 (3$\sigma$)} & \multicolumn{2}{c}{$>$28.85 (3$\sigma$)} \\
ACS F814W  & 27.25 &  0.05 & 27.28 &  0.04 & 28.44 &  0.12 \\
WFC3 105W  & 27.35 &  0.06 & 27.11 &  0.03 & 27.95 &  0.05 \\
WFC3 125W  & 27.43 &  0.08 & 27.02 &  0.03 & 27.86 &  0.06 \\
WFC3 140W  & 27.58 &  0.07 & 27.16 &  0.04 & 28.10 &  0.07 \\
WFC3 160W  & 27.55 &  0.07 & 27.29 &  0.04 & 28.07 &  0.09 \\
SHARDS F823W17 & 24.73 &  0.05 & 24.68 &  0.04 & 26.01 &  0.16 \\
\enddata
\end{deluxetable}

\begin{deluxetable}{lcc}
\centering
\tabletypesize{\small}
\tablewidth{0pc}
\tablecolumns{10}
\centering
\tablecaption{Summary of observed properties\label{table:physprop}}
\tablehead{\colhead{} & \colhead{A370-L57} &  \colhead{B2}}
\startdata
redshift ($z_{\textrm{Ly}\alpha}$) & 5.746 & 5.746 \\
rest-frame UV luminosity (M$_{UV}$)\tablenotemark{*} & -16.45$\pm$0.10 & -16.00$\pm$0.05 \\
UV spectral index ($\beta$) & -2.4$\pm$0.1 & -2.3$\pm$0.2 \\
half-light radius ($r_e$, pc)\tablenotemark{*} & 50--60 & - \\
Ly$\alpha$ luminosity (L(Ly$\alpha$)$_{obs}$, erg s$^{-1}$)\tablenotemark{*} & 7.7$\pm$0.4$\times$10$^{41}$ & 2.6$\pm$0.4$\times$10$^{41}$ \\
rest-frame Ly$\alpha$ equivalent width (EW(Ly$\alpha$), \AA) & 300--600 & 150--640 \\
\enddata
\tablenotetext{*}{Corrected for magnification using the lens model of \citet{Lagattuta17}}
\end{deluxetable}

\begin{deluxetable}{lccr@{$\pm$}l}
\centering
\tabletypesize{\small}
\tablewidth{0pc}
\tablecolumns{4}
\centering
\tablecaption{Stellar population parameters from SED-fitting\label{table:sedfit}}
\tablehead{\colhead{} & \colhead{range explored} & \colhead{best-fit} & \multicolumn{2}{c}{bayesian\tablenotemark{*}} }
\startdata
results for A$_V$ free & & \multicolumn{2}{c}{}\\
\hline
          log (t/yr)        & 6.0 -- 9.0 &  6.30 &  6.7 &  0.2\\
          log ($\tau$/yr)   & 6.0 -- 9.0 &  6.00 &  7.7 &  0.9\\
      Z                 & 0.0001 -- 0.02 & 0.0004 & 0.001 & 0.002\\
        log U             & -4.0 -- -1.0 & -1.00 & -2.5 &  1.0\\
          $f_{esc}^{LyC}$   & 0.0 -- 0.5 &  0.00 &  0.09 &  0.06\\
          A$_V$/mag         & 0.0 -- 1.0 &  0.45 &  0.32 &  0.17\\
              log (M$_*$/M$_\odot$) & -  &  6.66 &  6.5 &  0.2\\
        log (SFR/M$_\odot$yr$^{-1}$) & - &  0.66 &  0.1 &  0.3\\      
\\
results for A$_V$ = 0 & & \multicolumn{2}{c}{}\\
\hline
          log (t/yr)        & 6.0 -- 9.0 &  6.60 &  6.7 &  0.3\\
          log ($\tau$/yr)   & 6.0 -- 9.0 &  6.00 &  7.7 &  0.9\\
      Z                 & 0.0001 -- 0.02 & 0.0004 & 0.003 & 0.003\\
        log U             & -4.0 -- -1.0 & -1.50 & -2.4 &  1.0\\
          $f_{esc}^{LyC}$   & 0.0 -- 0.5 &  0.00 &  0.09 &  0.06\\
              log (M$_*$/M$_\odot$) & -  &  6.14 &  6.23 &  0.05\\
        log (SFR/M$_\odot$yr$^{-1}$) & - & -0.58 & -0.2 &  0.3\\

\enddata
\tablenotetext{*}{expectation value and 1-$\sigma$ uncertainty.}
\end{deluxetable}

\end{document}